\documentclass[iop]{emulateapj}
\usepackage{apjfonts}
\usepackage{graphicx}
\def\Fref#1{Figure~\ref{fig:#1}}
\def\Sref#1{Section~\ref{sec:#1}}
\def\Tref#1{Table~\ref{tab:#1}}
\def\Eref#1{Equation~(\ref{eq:#1})}
\def\KOI#1#2{KOI~#1#2}

\begin{document}

\shorttitle{Doppler boosting in Kepler light curves}
\shortauthors{van Kerkwijk et al.}
\submitted{Submitted to ApJ}
\title{Observations of Doppler boosting in Kepler light curves}

\author{Marten H. van Kerkwijk\altaffilmark{1,2}, 
        Saul A. Rappaport\altaffilmark{3}, 
        Ren\'e P. Breton\altaffilmark{1,2}, 
        Stephen Justham\altaffilmark{1}, 
        Philipp Podsiadlowski\altaffilmark{4}, 
        Zhanwen Han\altaffilmark{5}}

\altaffiltext{1}{Kavli Institute for Astronomy and Astrophysics,
  Peking University, Beijing, China} 
\altaffiltext{2}{Department of Astronomy and Astrophysics, University
  of Toronto, 50 St. George Street, Toronto, ON M5S 3H4, Canada; 
  \email{mhvk@astro.utoronto.edu}}
\altaffiltext{3}{37-602B, MIT Department of Physics and Kavli
  Institute for Astrophysics and Space Research, 70 Vassar St.,
  Cambridge, MA 02139, USA} 
\altaffiltext{4}{Sub-department of Astrophysics, University of Oxford,
  Keble Road, Oxford OX1 3RH, UK} 
\altaffiltext{5}{National Astronomical Observatories / Yunnan Observatory
  and Key Laboratory of the Structure and Evolution of Celestial
  Objects, Chinese Academy of Sciences, Kunming, 650011, China}

\defcitealias{rowe+10}{R10}
\defcitealias{rowe+10v1}{R10v1}

\begin{abstract}
  Among the initial results from {\em Kepler} were two striking
  light curves, for \KOI74 and \KOI81, in which the relative depths of
  the primary and secondary eclipses showed that the more compact,
  less luminous object was hotter than its stellar host.  That result
  became particularly intriguing because a substellar mass had been
  derived for the secondary in \KOI74, which would make the high
  temperature challenging to explain; in \KOI81, the mass range for
  the companion was also reported to be consistent with a substellar
  object.  We re-analyze the {\em Kepler} data and demonstrate that
  both companions are likely to be white dwarfs.  We also find that
  the photometric data for \KOI74 show a modulation in brightness as
  the more luminous star orbits, due to Doppler boosting.  The
  magnitude of the effect is sufficiently large that we can use it to
  infer a radial velocity amplitude accurate to $1{\rm\,km\,s^{-1}}$.
  As far as we are aware, this is the first time a radial-velocity
  curve has been measured photometrically.  Combining our velocity
  amplitude with the inclination and primary mass derived from the
  eclipses and primary spectral type, we infer a secondary mass of
  $0.22\pm0.03\,M_\odot$.  We use our estimates to consider the likely
  evolutionary paths and mass-transfer episodes of these binary
  systems.
\end{abstract}


\section{Introduction}
\label{sec:intro}

It has long been realized that precision photometry from space could
allow detailed probes of stellar interiors through pulsations, and of
planetary and stellar companions through transits, eclipses, and flux
variations due to tides and irradiation.  With the launch of {\em
  Kepler}, measurements with long-term precision of at least 1 part in
$10^5$ are now being collected for an unprecedented sample of stars
\citep{kepler1, kepler2}.

Among {\em Kepler}'s initial discoveries are two ``objects of
interest,'' \KOI74 and \KOI81, that have striking light curves, showing
both primary and secondary eclipses (\citealt{rowe+10}, hereafter
\citetalias{rowe+10}; where needed, we refer to the original version
posted on arXiv as \citetalias{rowe+10v1}).  In both systems, the
relative eclipse depths show that the more compact, less luminous
object is hotter than its stellar host.  This result became puzzling
as \citetalias{rowe+10v1} derived a clearly substellar mass for the
secondary in \KOI74; their mass range for the companion in \KOI81 was
also consistent with a substellar object.  If the smaller, hotter
companions were indeed planets or brown dwarfs, then it would not be
obvious how their properties could reasonably be explained.

In this work, we re-analyze the \emph{Kepler} data for \KOI74 and
\KOI81, and show that the photometric data for \KOI74 contain clear
evidence for Doppler boosting with orbital phase, in addition to
ellipsoidal light variations.  We use this to make what is, as far as
we are aware, the first measurement of a radial-velocity amplitude
from photometry.  With our amplitude, it is clear that the companion
is not substellar, but rather has a mass of $\sim\!0.2\,M_\odot$.  For
\KOI81, we find that the mass is likely similar.  We demonstrate that
all properties are consistent with the companions being low-mass white
dwarfs (WDs), and discuss possible evolutionary histories for these
systems.

\begin{figure*}
\centerline{\hfill
\hspace*{0.025\hsize}
\includegraphics[width=0.45\hsize,clip=y]{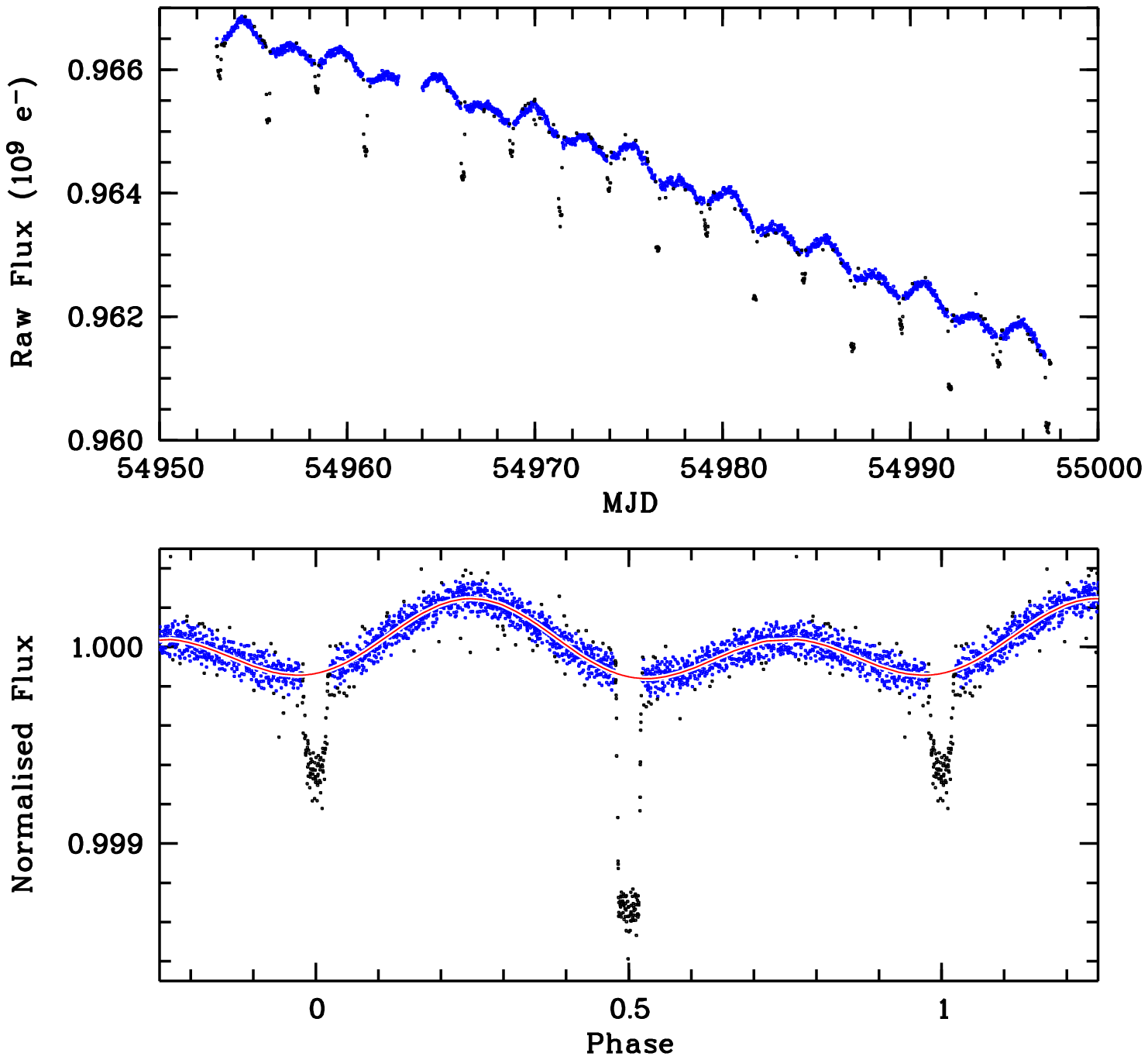}%
\hspace*{0.025\hsize}
\includegraphics[width=0.45\hsize,clip=y]{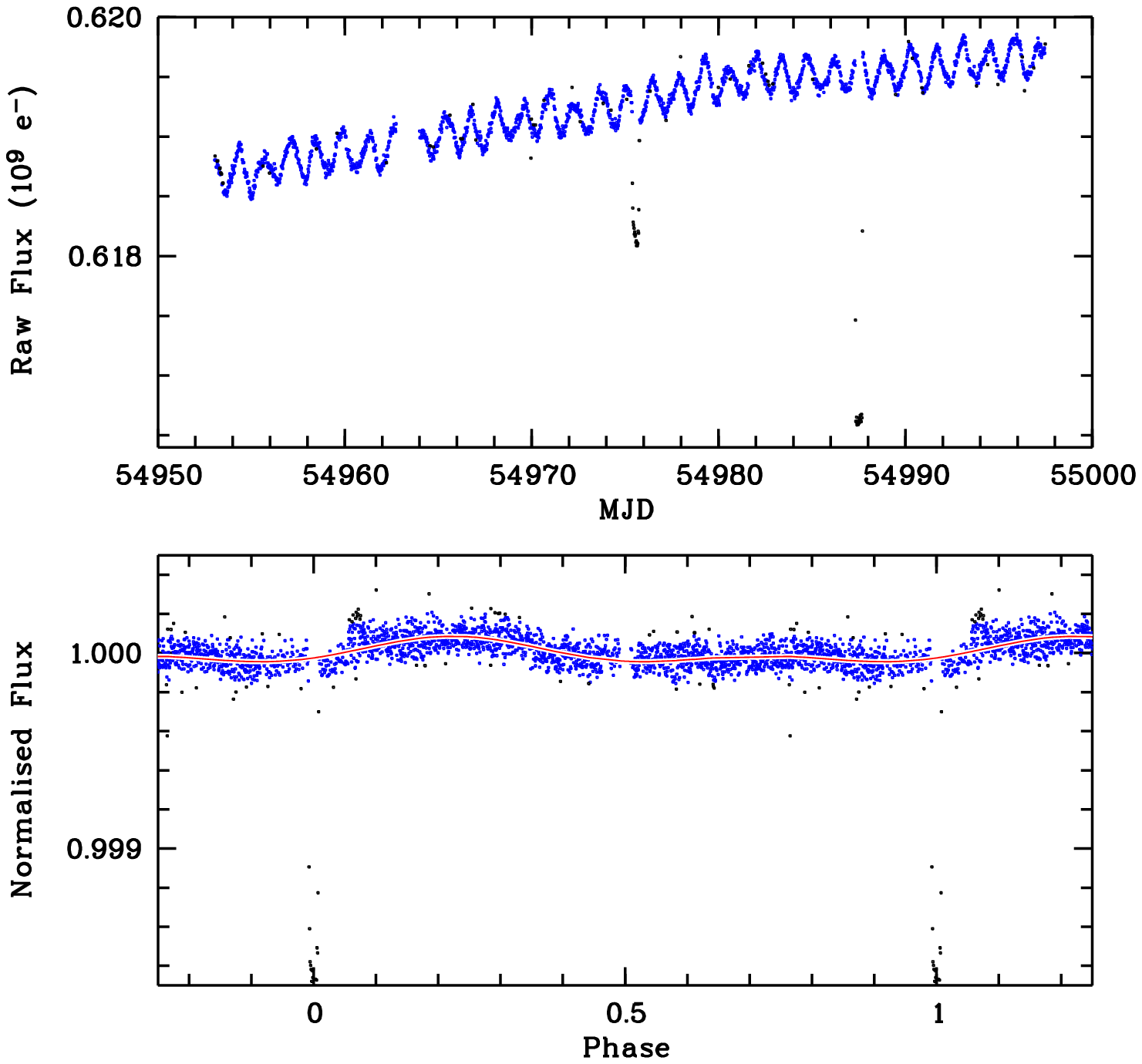}%
\hspace*{0.025\hsize}
\hfill}
\caption{{\em Kepler} light curves for \KOI74 (left) and \KOI81
  (right).  The top panels show the raw photometric data, and the
  bottom panels the detrended and folded light curves.  The larger dips
  correspond to eclipses of the hot companion, while the smaller ones
  correspond to transits over its parent star.  The smooth curve
  represents a fit to the blue data points -- those taken outside of
  transits and eclipses, and within $3\sigma$.  The fit includes two
  sine waves, at the orbital and half-orbital period, representing Doppler
  boosting due to orbital motion and ellipsoidal light variations,
  respectively. 
  \label{fig:lc}}
\end{figure*}

\section{The Light Curves Revisited}
\label{sec:reanalysis}
We retrieved the light curves of \KOI74 and \KOI81 from the archive,
and show the raw aperture fluxes in \Fref{lc}.  For both sources, one
sees the eclipses and transits superposed on roughly quadratic trends;
in \KOI81, one also sees the pulsations reported by
\citetalias{rowe+10}.  We first review what one can infer from the
transits and eclipses, and then discuss additional information seen at
other phases.

\subsection{Constraints from the Transit and Eclipse Light Curves}
\label{sec:eclipses}
The transit and eclipse light curves were fitted with detailed models
by \citetalias{rowe+10}, and we use their ephemerides and inclinations
(see \Tref{properties}).  For our purposes below, however, it is
useful also to have the ratio $R_1/a$, where $R_1$ is the radius of
the primary, and $a$ the semi-major axis.  This ratio is well
constrained by the eclipse, depending on the eclipse duration $t_{\rm
  e}$ (at half maximum depth) and inclination $i$ via $\sin^2(\pi
t_{\rm e}/P)\sin^2i=(R_1/a)^2-\cos^2i$ \citep{russ12}.  We measured
$t_{\rm e}/P$ graphically and list the values, as well as the inferred
ratios $R_1/a$ in \Tref{properties}.  Note that the derivation assumes
that the stars are spherical; if they are rotating rapidly, then the
constraint is mostly on the equatorial radius (see \Sref{rotation}).

The measurement of $R_1/a$, combined with Kepler's law, constrains the
mean density of the primary,
$\overline\rho_1=(R_1/a)^{-3}(3\pi/GP^2[1+q])$.  For both \KOI74
and \KOI81, the mass ratios $q=M_2/M_1$ are small, and, as discussed
by \citetalias{rowe+10}, the densities inferred for $q\ll1$, of
$\sim\!0.45$ and $0.17{\rm\,g\,cm^{-3}}$, respectively, are consistent
with those expected for main-sequence stars with masses of $2.2$ and
$2.7\,M_\odot$, respectively, as inferred from their spectral types
(see \Tref{properties}).

The ratio of the transit and eclipse depths $\epsilon_{\rm t}$ and
$\epsilon_{\rm e}$ equals the ratio of the surface brightnesses.
Thus, given the temperatures for the primaries inferred from their
spectra, one can estimate the temperatures of the secondaries.
Assuming blackbody emission, one has,
\begin{equation}
\frac{\epsilon_{\rm e}}{\epsilon_{\rm t}}
\simeq \frac{{\rm e}^{hc/\lambda kT_1}-1}{{\rm e}^{hc/\lambda kT_2}-1}.
\label{eq:sbratio}
\end{equation}
Using the numbers from \Tref{properties} and assuming an effective
wavelength $\langle\lambda\rangle=6000\,$\AA, we infer $T_2=13000$ and
$17000\,$K for \KOI74 and \KOI81, respectively, with uncertainties
somewhere between 5\% and~10\%.  These temperatures are higher than
inferred by \citetalias{rowe+10}, who assumed that the eclipses and
transits measured bolometric surface brightness.  (We note that, of
course, one could derive more precise values by folding blackbodies
or model atmospheres through the {\em Kepler} passband, but this would not
necessarily be more accurate as long as one does not account properly
for, e.g., gravity darkening on what may well be a rapidly rotating
primary.  In any case, for our purposes the above estimates suffice.)

\subsection{Ellipsoidal Variations and Doppler boosting in \KOI74}
\label{sec:doppler}

For \KOI74, in addition to the long-term trend, eclipses, and
transits, the raw fluxes show sinusoidal modulations at the orbital
and half the orbital period (\Fref{lc}).  We fitted the data outside
the eclipses and transits with a 6th-degree polynomial and sine waves
at the orbital frequency and its harmonic (and iterated once,
rejecting $>\!3\sigma$ outliers).  In the bottom panel of the figure,
where fluxes normalized by the polynomial component of the fit are
shown as a function of phase, the modulations are very obvious.  They
have much larger amplitude than is seen in Fig.~1 of
\citetalias{rowe+10v1}, likely because these authors originally had
removed most of the signal in their pre-display median filtering;
their later version shows a larger signal, more similar to what we
find.  (Related, we note that the two archive light curves include a
`corrected' flux.  In the second, longer light curve, the signal is
clearly present, but in the first, shorter one, it is not, apparently
having been removed by the `correction' applied in the {\em Kepler}
pipeline.)

From our fit, we infer fractional amplitudes
$\mathcal{A}_1=(1.082\pm0.013)\times10^{-4}$ and
$\mathcal{A}_2=(1.426\pm0.018)\times10^{-4}$ for the fundamental and
the harmonic, respectively.  As can be seen in \Fref{lc}, these two
fit the data quite well; although the fit is not formally acceptable
($\chi^2_{\rm red}=1.5$ for 1843 degrees of freedom), the residuals
appear quite white.  A Fourier transform of the residuals confirms
this impression; there is no power above fractional amplitude
$10^{-5}$, apart from one faint signal with
$\mathcal{A}\simeq1.6\times10^{-5}$ and $f\simeq1.66{\rm\,d^{-1}}$.
In \Fref{lc}, one sees that the maximum of the fundamental coincides
with one of the maxima of the harmonic; our fit gives a phase
difference of $-0.009\pm0.003$ cycles (of the orbital period; the
maximum of the fundamental occurring slightly earlier).  The
fundamental has essentially the same offset ($-0.008\pm0.002$ cycles)
from the descending node of the massive star, while the harmonic has
maxima coincident with the nodes to within the measurement errors
($0.0005\pm0.0008$ cycles).

The above precisions are remarkable, and hence we checked whether the
error estimates were reliable.  We found that the amplitudes do depend
on how we model the long-term trend; if we reduce the polynomial to
4th degree, we find $\mathcal{A}_{1,2}=(1.034,1.430)\times10^{-4}$;
if we fit the pipeline `corrected' fluxes in the longer light curve
with a constant and the two sine waves, we obtain
$\mathcal{A}_{1,2}=(1.076,1.515)\times10^{-4}$.  This suggests that
the true uncertainty in the amplitudes is about 5\%, although we note
that these latter fits are not as good (we find $\chi^2_{\rm
  red}=1.7$, and the phases do not agree as well with expectations).
Obviously, the uncertainties will reduce substantially as further data
are added.

Both signals have natural if, for the fundamental, unusual
interpretations.  We believe that the fundamental is due to orbital
Doppler boosting, an effect which, as far as we are aware, was first
detected and discussed by \citet{maxtmn00}; it is discussed in the
context of {\em Kepler} by \citet{loebg03} and \citet{zuckma07}.
Here, the precision of the {\em Kepler} data allows us to go beyond
detection, and use the Doppler boosting to measure a star's
radial-velocity amplitude.

Briefly, Doppler boosting occurs because as the primary orbits, its
spectrum is Doppler shifted, its photon emission rate is modulated,
and its emission is slightly beamed forward.  Following
\citet{loebg03}, we use the relativistic invariant quantity
$I_\nu/\nu^3$ (or $I_\lambda\lambda^5$), where $I_\nu$ is the specific
intensity, and $\nu$ the frequency.  Then, for a radial velocity
$v_r$, the photon rate $n_\gamma$ observed by a telescope with given
effective area $A_\lambda$ is given by,
\begin{equation}
n_\gamma = \int_\lambda A_\lambda\frac{F_\lambda}{hc/\lambda}{\rm\;d}\lambda
  = \int_\lambda A_\lambda\frac{F_{\lambda^\prime}(1+v_r/c)^{-5}}{hc/\lambda}{\rm\;d}\lambda,
\end{equation}
where $F_\lambda\equiv \int_\Omega I_\lambda{\rm\,d}\Omega$ is the observed flux
($\Omega$ is the solid angle subtended by the star; note that
the integral is correct only to first order in $v_r/c$).  For a narrow
band, the fractional variation would then be
\begin{equation}
\frac{\Delta n_\gamma}{n_\gamma} \equiv f_{\rm DB}\frac{v_r}{c} =
\left(-5-\frac{{\rm d}\,\ln F_\lambda}{{\rm d}\,\ln\lambda}\right)\frac{v_r}{c}
\simeq-\frac{(hc/\lambda kT){\rm e}^{hc/\lambda kT}}{{\rm e}^{hc/\lambda kT}-1}\frac{v_r}{c},
\end{equation}
where we introduced a Doppler boost pre-factor $f_{\rm DB}$ and where
the approximate equality is for blackbody emission.  For \KOI74
($T=9400\,$K), observed with {\em Kepler}
($\langle\lambda\rangle\simeq6000\,$\AA), one thus expects a signal
with fractional amplitude of $\sim\!2.8(K/c)$, where $K$ is the
radial-velocity amplitude.  For a more precise estimate, we folded a
$9500\,$K model atmosphere \citep{muna+05} through the {\em Kepler}
response.  This yields $f_{\rm DB}=2.25$; the difference with the
blackbody results from the fact that the model continuum is closer to
Rayleigh Jeans inside the {\em Kepler} bandpass, mimicking the
emission from a hotter blackbody.  In this temperature range, the
boost pre-factor scales approximately linearly with temperature; for
$T=9000$ and 10000\,K, we find $f_{\rm DB}=2.37$ and 2.15,
respectively.  Including reddening of $E(B-V)=0.15$ (from the {\em
  Kepler} Input Catalog) has only a small effect, yielding $f_{\rm
  DB}=2.21$.  Given that adding the extinction is equivalent to a
relatively large change in effective area, the small change in $f_{\rm
  DB}$ found also means that uncertainties in the response are not
important.

Given the above, using $f_{\rm DB}=2.21$, our observed light amplitude
corresponds to a radial-velocity amplitude of $14.7\pm1.0~{\rm
  km\,s^{-1}}$, where we assumed 5\% uncertainties in both
$\mathcal{A}_1$ and in $f_{\rm DB}$.  The corresponding mass function
is $\mathcal{M}=0.0017\pm0.0004\,M_\odot$.  If we adopt a mass for the
primary star of $2.22\,M_\odot$ (\citetalias{rowe+10}; see
\Sref{eclipses}), this yields $M_2\simeq0.22\,M_\odot$.

Turning now to the harmonic, as in \citetalias{rowe+10}, we interpret
it as changes in the observable surface area due to tidal distortion.
The expected amplitude is given by,
\begin{equation}
\mathcal{A}_2 = f_{\rm EV}\frac{M_2}{M_1}\left(\frac{R_1}{a}\right)^3\sin^3 i,
\label{eq:ellipsoidal}
\end{equation}
where $f_{\rm EV}$ is a pre-factor of order unity that depends on the
limb darkening and gravity darkening.  For our system, the eclipse and
transit light curves yield $i$ and $R_1/a$ (\Sref{eclipses}).
Combining these with the amplitude yields a mass ratio
$M_2/M_1=0.092/f_{\rm EV}$.  If we again adopt a primary mass of
$2.22\,M_\odot$, and take $f_{\rm EV}\simeq1.5$ (see
\Sref{numerical}), we derive $M_2\simeq0.14\,M_\odot$, somewhat below
the value obtained above using the radial velocity amplitude, but
roughly consistent, given the probably larger uncertainties associated
with determining the mass ratio via ellipsoidal light variations (see
\Sref{rotation}).

\begin{deluxetable*}{lcccc}
\tablewidth{0.75\textwidth}
\tablecaption{\label{tab:properties} System parameters} 
\tablehead{  & \multicolumn{2}{c}{\dotfill \KOI74\dotfill} &
  \multicolumn{2}{c}{\dotfill \KOI81\dotfill}\\
\colhead{Property} & \colhead{Primary} & \colhead{Secondary} & \colhead{Primary} & \colhead{Secondary}  }
\startdata
$P_{\rm orb}$ (d)\dotfill &  \multicolumn{2}{c}{$5.18875\pm0.00008$}&  \multicolumn{2}{c}{$23.8776\pm0.0020$}  \\
$i~({}^\circ)$\dotfill & \multicolumn{2}{c}{$88.8\pm0.5$}& \multicolumn{2}{c}{$88.2\pm0.3$}\\
Eclipse duration (cycle)\dotfill & \multicolumn{2}{c}{$0.0362\pm0.0004$} & \multicolumn{2}{c}{$0.058\pm0.004$}\\
Eclipse depth ($10^{-5}$)\dotfill & $51\pm5$ & $118\pm5$ & $160\pm5$ & $496\pm5$ \\
$R/a$ \dotfill & $0.116\pm0.002$ & $0.0026\pm0.0002$ & $0.058\pm0.004$ & $0.0023\pm0.0003$ \\
Spectral type\dotfill & A1\,V&\nodata&B9--A0\,V&\nodata\\
Radius ($R_\odot$)\dotfill & $1.90^{+0.04}_{-0.05}$ & $0.043\pm0.004$  & $2.93\pm0.14$ & $0.117\pm0.012$ \\
$T_{\rm eff}$ (K)\dotfill &  $9400\pm150$ & $13000\pm1000$  & $10000\pm150$ & $17000\pm1300$ \\
Luminosity ($L_\odot$)\dotfill &  $25.6\pm2.4$ & $0.05\pm0.02$ & $77.3\pm9.6$ & $0.9\pm0.4$ \\
Velocity amplitude $({\rm\,km\,s^{-1}})\ldots$& $14.7\pm1.0$&\nodata & $\sim\!7$& \nodata\\
Mass ($M_\odot$)\dotfill & $2.22^{+0.10}_{-0.14}$ & $0.22\pm0.03$ & $2.71^{+0.19}_{-0.11}$ & $\sim\!0.3$ \\
Model mass ($M_\odot$)\dotfill & \nodata & $0.20\pm0.03$ & \nodata & $0.25\pm0.03$
\enddata
\tablecomments{The periods and inclinations are taken from
  \citet{rowe+10}, as are the spectral types of the primaries and the
  quantities inferred from those.  Systematic errors are possible if
  the star has accreted significant matter, or is rotating rapidly.
  The eclipse durations and depths were measured graphically, and have
  ``chi-by-eye'' uncertainties.  For the temperatures of the
  secondaries, we assumed an 8\% systematic uncertainty in our
  conversion of surface brightness to temperature.  The velocity
  amplitude is measured from the Doppler boost signal, and its
  uncertainty includes possible systematic effects related to the
  detrending of the data and the precise spectrum of the primary. The
  model mass is based on the theoretical relation between orbital
  period and WD mass (see \Eref{McP}).}
\end{deluxetable*}

\subsection{Other Signals}

Apart from the signals discussed above, one might also expect to see
contributions of other factors, in particular the Doppler boosting and
ellipsoidal variations of the secondary, irradiation effects on both
primary and secondary, and possibly influences from an eccentric
orbit.  Following the procedure above, we find that the fractional
amplitudes for the fundamental and harmonic of the secondary, relative
to the secondary's flux, are $\sim\!10^{-3}$ and
$\sim\!2\times10^{-7}$, respectively (where we used $K_2\simeq150~{\rm
  km\,s^{-1}}$, $f_{\rm DB}=1.9$, $R_2/a\simeq0.0026$, and $f_{\rm
  EV}=1$).  Since the secondary's flux is only 0.2\% of the total
flux, however, the contribution to the observed signal is negligible.

For irradiation of the primary by the secondary, one expects a change
in luminosity of $\Delta L_1\simeq\pi R_1^2 (L_2/4\pi a^2)$,
corresponding to a fractional amplitude $\frac{1}{2}\Delta
L_1/L=\frac{1}{2}(L_2/L)(R_1/2a)^2\simeq3.0\times10^{-6}$ (where
$L\simeq L_1$ is the total luminosity).  Within the {\em Kepler} passband,
this should lead to a modulation of $\sim\!1.7\times10^{-6}$.
Similarly, for the secondary, one expects a fractional amplitude of
$\frac{1}{2}(R_2/2a)^2\simeq0.8\times10^{-6}$ in bolometric flux, and
a signal of $\sim\!0.4\times10^{-6}$ in the {\em Kepler} passband.  The two
will have opposite phase; thus, the net expected signal has an
amplitude of $\sim\!1.3\times10^{-6}$, and is phased such that maximum
occurs at the time of transit.  This signal may be responsible for the
fact that the fundamental is slightly out of phase with what is
expected from Doppler boosting; the observed out-of-phase amplitude is
$\mathcal{A}_1\sin\Delta\phi_1=(9\pm3)\times10^{-7}$, consistent with
what we derived above.

Finally, we consider possible signals from a non-circular orbit.  For
small eccentricity, both the Doppler boost curve and the ellipsoidal
variations would no longer be pure sinusoids, but have some
contribution from higher harmonics.  For the easier case of the
Doppler boost curve, we can constrain the eccentricity from the
absence of a signal at the first harmonic that is out of phase with
that expected from ellipsoidal variations (to a limit of
$\Delta\phi_2=0.0005\pm0.0008$ cycles; \Sref{doppler}).  We find that,
at 95\% confidence,
$e\sin\omega=(\mathcal{A}_2/\mathcal{A}_1)\sin4\pi\Delta\phi_2<0.03$
(where $\omega$ is the longitude of periastron, and this expression is valid
in the limit of small $e$).  One sees that one cannot constrain the
case where periastron occurs at one of the nodes, since in that case
the harmonic would be hidden by the ellipsoidal variations.  A
complementary constraint, however, comes from the fact that the
mid-transit and mid-eclipse times are consistent with a circular
orbit.  The measured offset from half a cycle is
$\Delta\phi=0.0003\pm0.0008$ cycles, which, at 95\% confidence,
implies $e\cos\omega=\pi\Delta\phi<0.006$ (with the expression again
being valid in the limit of small $e$).  Thus, from both constraints
combined, we conclude that the orbit is circular to within $e<0.03$.

\subsection{Application to \KOI81}
\label{sec:koi81}
For \KOI81, a detailed analysis is not possible, since variations on
its much longer orbital period are strongly covariant with the trends
in the raw data.  Furthermore, the star is clearly pulsating; a
Fourier transform shows that there is signal at numerous frequencies.
In order to look for orbital modulation, we fit the raw fluxes with
orbital and half-orbital modulations, a 3rd degree polynomial, as well
as sine waves at the five largest pulsational signals, at 0.362,
0.723, 0.962, 1.32, and $2.08{\rm\,d^{-1}}$ (some of these are
harmonically related, as is often the case for multi-periodic
pulsators).  From the fit, orbital modulation does appear to be
present, as is also clear in the folded, normalized light curve, where
we divided the fluxes by the fitted trend and pulsational signals (see
\Fref{lc}).  Our fit yields amplitudes
$\mathcal{A}_1\simeq5\times10^{-5}$ and
$\mathcal{A}_2\simeq4\times10^{-5}$, which corresponds to
$K_1\simeq7{\rm\,km\,s^{-1}}$ and $q\simeq0.2/f_{\rm EV}$.  Because of
the long orbital period, the results are very sensitive to, e.g., the
degree of the polynomial used.  Comparing with a fit using a 6th
degree polynomial, as well as one using the `corrected' flux from the
{\em Kepler} archive light curve (the latter for the second, longer
light curve only), the amplitude of the fundamental changes by up to a
factor of 2, while that of the harmonic changes by $\sim\!20$\%.
Furthermore, we find relatively large phase offsets, of $\sim\!0.03$
cycles.  Given this, we will not try to infer quantities, but simply
note that for a $2.7\,M_\odot$ primary with a $0.3\,M_\odot$
companion, the predicted amplitudes are $\sim\!8\times10^{-5}$ and
$3\times10^{-5}$, respectively, consistent with the observations.

\subsection{Verification Using a Light Curve Synthesis Code}
\label{sec:numerical}
We tried to verify our semi-analytical estimates above using a
light-curve synthesis code, which is similar to that of
\citet{orosh00}, and has been used previously to model irradiated
pulsar companions \citep{stap+99}.  The code accounts for tidal
distortion and stellar rotation in the Roche approximation, includes
irradiation and gravity darkening, and calculates the flux using
NextGen model atmospheres \citep*{hausab99}, integrated over the
{\em Kepler} passband.  It cannot yet deal with eclipses, and hence we only
attempt to reproduce the orbital modulations.

As input parameters, we chose to use the set $P_{\rm orb}$, $t_0$,
$i$, $T_1$, $R_1/a$, $K_1$, $P_{\rm1,rot}$, $q$, and $T_{\rm irr}$,
where the ones not mentioned before are $t_0$, the time of conjunction,
and $T_{\rm irr}$, the effective temperature corresponding to the
secondary flux absorbed by the primary (i.e., $\sigma T_{\rm
  irr}^4=(1-A)L_2/4\pi a^2$, with the albedo $A$ expected to be close
to zero for a radiative star).  Our main goal is to constrain $q$,
$K$, and $T_{\rm irr}$; we assume the other parameters are as inferred
from the eclipses.  For our fits, we also assume that the star is
corotating with the orbit ($P_{\rm1,rot}=P_{\rm orb}$); we will return
to this in \Sref{rotation}.  We add to the model the tiny flux
contribution from the secondary (taken to be constant at the level
indicated by the eclipses).

Fitting the raw data of both sources to our model, we confirmed that
for \KOI81, the results depend sensitively on how one fits the
long-term trend.  We thus decided to focus on \KOI74.  We searched
over a grid in $q$, $K_1$, and $T_{\rm irr}$, and fitted for the best
4th-degree polynomial at each position.  The best-fit synthetic
light curve is nearly indistinguishable from the two-sine fit, and has
the parameters $q=0.0689\pm0.0009$, $K=14.2\pm0.2{\rm\,km\,s^{-1}}$,
and $T_{\rm irr}=760\pm40\,$K.  We did not exclude outliers, and
therefore our fit is poorer than the analytical one ($\chi^2_{\rm
  red}=2.6$); we scaled the uncertainties such that $\chi^2_{\rm
  red}=1$ (but did not attempt to include uncertainties in effective
temperature, etc.).

These results reproduce our analytical estimates.  The offset in $K_1$
would be even smaller if we had included reddening in our numerical
model (as this affects $f_{\rm DB}$; see \Sref{doppler}).  The mass
ratio matches our estimates for $f_{\rm EV}=1.34$, quite close to the
value of $f_{\rm EV}=1.63$ inferred from the tables of \citet{beec89}.
The secondary luminosity inferred from the irradiation is
$0.08\pm0.02\,L_\odot$ (for orbital separation $a=16.4\,R_\odot$ and
$A=0$, i.e., assuming complete absorption and reradiation), consistent
with our estimate from the surface brightness ratio.  Below, we give
our best estimates of the masses based on these results, but first we
discuss possible uncertainties due to rapid rotation.

\subsection{Uncertainties due to Rotation}
\label{sec:rotation}
In many ways, the light curves of \KOI74 and \KOI81 are easy to model,
since all effects are small and can thus adequately be treated as
perturbations.  The one parameter that we have not yet constrained,
however, is the rotation of the primary.  Since rapid rotation is
expected in the context of the evolutionary scenario described below,
we discuss the observational consequences in some detail.  

Rotation of the primary has a number of effects.  First, the
constraints on inclination and radius of the primary from the eclipse
and transit change: the true inclination $i^\prime$ will be closer to
$90^\circ$ than the inclination $i$ inferred assuming a spherical
star, the radius will be smaller, and the radius that is constrained
is the equatorial radius $R_{\rm eq}$ (assuming aligned rotation).
Furthermore, the estimate of the mean density increases, to
$\overline\rho_1^\prime\simeq(R_{\rm eq}/a)^{-3}(3\pi/GP^2[1+q])(R_{\rm
  eq}/R_{\rm pole})$, where $R_{\rm pole}$ is the polar radius; thus,
one would infer a somewhat smaller primary mass.  For \KOI74, which
has an inclination very close to $90^\circ$, we conclude that it would
have $R_{\rm eq}/a\simeq R_1/a$.  For \KOI81, $R_{\rm eq}/a$ would be
somewhat smaller than $R_1/a$ inferred from the eclipses.

The second effect of a rapidly rotating primary is that its pole will
be hotter than the equator; hence, looking at the equator, one will
underestimate its temperature, and thus its luminosity and mass.  This
also affects the estimates for the secondary, since the ratios of the
radii ($R_2/R_1$), based on the transit depth, and that of the surface
brightnesses (\Eref{sbratio}) will be different from their true
values: the cooler equator will lead to a larger inferred $R_2/R_1$
and smaller inferred $T_2$.  For the companion radius $R_2$, however,
the change in $R_2/R_1$ is partly compensated by the fact that a
rapidly rotating primary also would have a smaller area
($R_1^\prime\simeq(R_{\rm eq}R_{\rm pole})^{1/2}$; see above).  We
have used our model code to test an illustrative case of extreme
rotation, with the primary rotation rate 20 times faster than the
orbital one (corresponding to $v\sin i\simeq370{\rm\,km\,s^{-1}}$),
and a ratio of equatorial to polar radius $R_{\rm eq}/R_{\rm
  pole}=1.32$ (similar to what is measured for Regulus; see
\citealt{mcal+05}).  Keeping the flux-weighted temperature at 9400\,K,
we find equatorial and polar temperatures of $\sim\!7800$ and
$11600\,$K, respectively.  For this particular example, we estimate
that the radius of the hot companion would be
$R_2^\prime\simeq0.045\,R_\odot$, i.e., increased by only 5\% from that
inferred for a slowly rotating primary, and its temperature
$T_2^\prime\simeq10500\,$K.

Rapid rotation also indirectly affects the ellipsoidal variations.
First, because of the different temperature distribution on a rapidly
rotating primary, different regions are weighted differently, and as a
result the amplitude of the ellipsoidal variations changes.  From our
model code, we find that for the above-mentioned case, the predicted
amplitude of the ellipsoidal variations increases by a
factor~$\sim\!2$.  The effect is strongly non-linear, however;
reducing the rotation rate from 20 to 15 times faster than orbital,
the amplitude increases only by a factor of~1.3. 

Another indirect effect of a rapidly rotating primary may be that the
inferred $q$ does not necessarily correspond to the true mass ratio.
An implicit assumption underlying the equilibrium tide is that the
primary corotates with the orbit.  For stars that do not rotate
synchronously, the work of, e.g., \citet*{pfahap08}, suggests that the
equilibrium tide may be a rather poor approximation to reality.  Thus,
it may be that one cannot reliably infer mass ratios from ellipsoidal
variations for systems that are not tidally locked.

We note that the light curves may contain a clue to the rotation rates.
In particular, for \KOI74, we found evidence for a weak modulation at
$\sim\!0.6\,$\,d, which is about 9 times faster than the orbital
period, and corresponds to 30\% of break-up for a $2.2\,M_\odot$ and
$1.9\,R_\odot$ star.  For \KOI81, it may be possible to infer the
rotation rate from the pulsations by looking for rotational splitting
(though for high rotation rates this is non-trivial).

Of course, it would be simpler and more reliable to measure the
rotational broadening spectroscopically, and the spectra mentioned by
\citetalias{rowe+10} may already hold the answer, but we do not have
access to those.  Nevertheless, we note in closing a much more subtle
consequence of rapid rotation, which is the photometric analog of the
Rossiter-McLaughlin effect \citep{ross24,mcla24}.  Assuming the
rotation is aligned with the orbit, at the start of the transit more
blueshifted light is being blocked and at the end more redshifted
light.  For a rotational velocity of $300{\rm\,km\,s^{-1}}$, Doppler
boosting induces a signal in the blocked light with a fractional
amplitude of $\sim\!2\times10^{-3}$.  This is diluted by the unblocked
light, i.e., by a factor equal to the transit depth.  Thus, for \KOI74
and \KOI81, one expects net signals of $\sim\!1\times10^{-5}$ and
$3\times10^{-6}$, respectively, which may be detectable by averaging
about 100 transits.

\subsection{Adopted Masses}
\label{sec:choices}
In principle, given our inferred values of $K_1$ and $q$, combined
with the inclination, one can derive both masses without further
assumptions.  In practice, however, these masses are very uncertain,
since the uncertainties in $K_1$ and $q$ enter to high powers (for
small $q$, $M_1\propto K_1^3q^{-3}$ and $M_2\propto K_1^3q^{-2}$).
Given that these uncertainties are of order 5\% for $K_1$ and likely
larger for $q$ (see above), the 1$\sigma$ fractional errors are
$\gtrsim\!20$\%, and the distributions are highly non-Gaussian.

Instead, we proceeded by assuming the primary has a mass of
$2.22^{+0.10}_{-0.14}\,M_\odot$, as inferred from its spectral type
and mean density (\citetalias{rowe+10}; see also \Sref{eclipses}).
Then, using $K_1$ and~$i$, we calculate a mass of
$0.22\pm0.03\,M_\odot$ for the secondary.  If instead we use the mass
ratio $q$, inferred from the ellipsoidal variations, we find
$M_2=0.15\,M_\odot$, with an uncertainty that is difficult to estimate
because we cannot be sure the star is in corotation (see above).  We
will proceed by using the value inferred from $K_1$ (see
\Tref{properties}).  For \KOI81, we do not have a good constraint on
$K_1$ (or $q$), but the signals that are present are consistent with a
similar secondary mass, of $\sim\!0.3\,M_\odot$.

\begin{figure}
\includegraphics[angle=270,width=0.95\hsize]{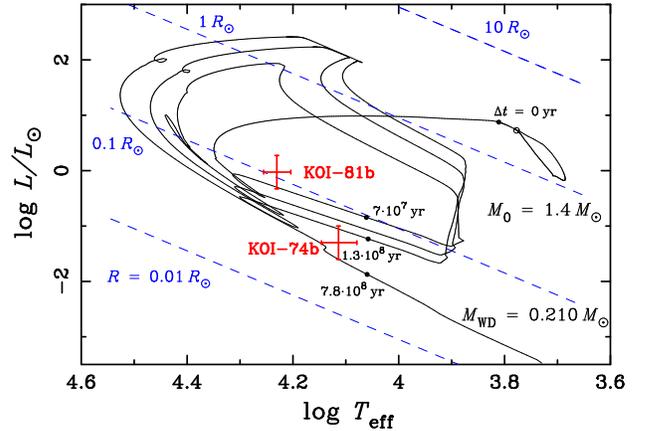}
\caption{\label{fig:cooling} Representative evolutionary track in the
  Hertzsprung-Russell (HR) diagram, illustrating the evolution of the
  progenitors of the hot companions of \KOI74 and \KOI81. The solid
  black curve shows the track of a star with an initial mass of
  $1.4\,M_\odot$ that starts to transfer mass to a companion star just
  after leaving the main sequence and ultimately becomes a
  $0.21\,M_\odot$ helium WD. Because of residual hydrogen in
  the envelope at the end of the main mass-transfer phase, the
  degenerate remnant experiences three hydrogen shell flashes,
  producing three large loops in the HR diagram (of progressively
  larger area), before it settles on a classical He WD cooling track.
  Selected ages since the end of the main mass-transfer phase
  (labelled as ``$\Delta t = 0\,{\rm yr}$'') are given next to filled
  circles. The dashed lines give lines of constant radius as
  indicated. The positions of \KOI74b and \KOI81b are also shown with
  conservative error bars. They are consistent with the expected
  location for $\sim\!0.2$--0.25\,$M_\odot$ He WDs either on
  a classical He WD cooling track or during one of the relatively
  long-lived phases of a H-flash loop.}
\end{figure}

\section{Evolutionary History}
\label{sec:evolution}

An obvious explanation for these hot, compact companions is that they
are WD stars. \citetalias{rowe+10} have discussed this
possibility for \KOI81, but their initial very low mass estimate for
the companion in \KOI74 ($\leq\!0.032\,M_\odot$;
\citetalias{rowe+10v1}) was inconsistent with a WD explanation.  Since
our reanalysis of the data gives the mass of the companion in \KOI74
as $\sim\!0.2\,M_\odot$, a WD companion is no longer excluded.
Cooling tracks from \citet{pane+07} for helium-core WDs of appropriate
mass are broadly consistent with the companion luminosities and
temperatures in Table~\ref{tab:properties} (somewhat more easily for
\KOI74 than for \KOI81).  Figure \ref{fig:cooling} compares a cooling
track from our own calculations\footnote{In the binary calculation
  shown in \Fref{cooling} (from the library of \citealt*{podsrp02}),
  the companion is a neutron star.  The tracks for the \KOI74 and
  \KOI81 systems are expected to be very similar (see also
  \citealt{rappph09}). For more details on the hydrogen shell flashes
  in these tracks, see \citet{podsrp02} and \citet*{nelsdm04}.}  for a
$\sim\!0.2\,M_\odot$ helium-core WD to the WDs in \KOI74 and 81.

Binary systems resembling these may well be common; the well-known
star Regulus ($\alpha$ Leonis) has recently been found to be a
spectroscopic binary with component masses $\sim\!0.3$ and
$3.4\,M_\odot$, and an orbital period of 40.11\,d \citep{gies+08}.
The lower-mass component likely is a WD, making Regulus and its
companion remarkably similar to the pair of systems considered here.
The evolution of Regulus has been considered in detail by
\citet*{rappph09}.

To estimate the occurrence more quantitatively, we note that {\em
  Kepler} observed only $\sim\!400$ stars with temperatures above
about 9000\,K \citep{bata10}.  Since the eclipse probabilities for
systems like \KOI74 and \KOI81 are $\sim\!12$\% and 6\%, respectively,
the detection of two such WD companions suggests that a surprisingly
large fraction of such stars must have WD companions.  With
only two detections, it is difficult to estimate the orbital period
distribution.  Taking an average eclipse probability of
$\mathcal{P}\simeq9$\%, then for a fraction $\mathcal{F}$ of A and B
stars with suitably close WD companions, the mean number of
expected eclipses among the {\em Kepler} sample would be
$\sim\!400\mathcal{FP}$.  The Poisson probability for finding two 
or more eclipses out of 400 systems is $\sim\!50$\% for
$\mathcal{F}=0.05$ and $\sim\!25$\% for $\mathcal{F}=0.03$.
Therefore, we conclude that 1 out of every 20 to 30 A and B stars is
in a close, few tens of days, binary with a WD companion.

The above abundance is very high, and, if confirmed, has significant
implications for our understanding of stellar and binary evolution
(see also \citealt{dist10}).  For instance, the descendants of these
systems should also be relatively numerous---and likely
interesting---stars.  When the currently more massive star leaves the
main sequence and expands to overfill its Roche lobe, the resulting
mass transfer will be dynamically unstable, and likely lead to a
merger (even more likely than for Regulus, which has a longer period;
see \citealt{rappph09}).  In this merger, the addition of a helium WD
to the helium core of the subgiant star could lead directly to core
helium ignition, bypassing the usual gradual buildup of the helium
core via hydrogen shell burning.  The merger product would also be
rapidly rotating, and hence the natural outcome would appear to be a
rapidly-rotating red clump or horizontal branch star (such as observed
by, e.g., \citealt{behr+00} and \citealt{behr03}).

Of course, while the above is interesting, it is possible that we are
dealing with a statistical fluke.  Indeed, if WD companions
are very common among A stars, it seems odd that none has been
reported so far for any of the much more numerous cooler stars in the
{\em Kepler} sample.  On the other hand, this may indicate that binary
systems with initially lower-mass stars, which would evolve to G and F
stars with low-mass WD companions, have a difficult time
forming and/or surviving.

Turning now to the formation of \KOI81 and \KOI74, for both the
initial masses of the WD progenitors must have been greater than the
initial masses of the current A stars, since the progenitors of the
WDs have evolved more rapidly.  Some form of binary interaction is
needed in order to remove mass and produce the WDs.  Either stable
mass-transfer from the more massive to the less massive star has
occurred; in this case, the current mass of the A star may be
substantially higher than its initial mass (see, e.g.,
\citealt{rappph09}).  Alternatively, the system may have passed
through a common-envelope phase \citep{pacz76}.

\begin{figure}
\includegraphics[width=0.925\hsize]{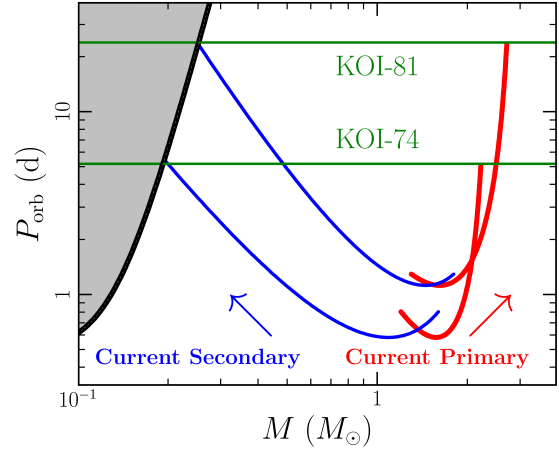}
\caption{\label{fig:histories} Example evolutionary histories in the
  mass-period plane. The red curves are for the mass-gaining stars,
  the blue curves for the mass-losing stars. The orbital periods of
  \KOI74 and \KOI81 are marked with green lines. The $M_{\rm
    wd}$--$P_{\rm orb}$ relation, where the mass transfer terminates,
  is marked by a black curve.  Many different such potential histories
  could be constructed.  }
\end{figure}

\subsection{Stable Mass Transfer}

If the mass loss from the progenitor of the WDs were simply through
stable Roche-lobe overflow, then we could assume that the current
orbital periods are very close to the orbital periods at the end of
mass transfer, i.e., the point at which the mass of the envelope about
each current WD became too small to maintain a giant structure and
collapsed.\footnote{A-type stars typically experience no magnetic
  braking, which might otherwise have shortened the orbital period
  since the end of mass-transfer.}  Hence we can apply the
relationship between core mass and radius to predict the mass of each
WD based on the current orbital periods.  Using this relationship,
\citet{rapp+95} approximated the orbital period $P_{\rm orb}$ as,
\begin{equation} 
\label{eq:McP}
P_{\rm orb} \approx 1.3 \times 10^{5} \frac{M_{\rm wd}^{6.25}}{(1+4M_{\rm wd}^{4})^{1.5}} ~~{\rm days},
\end{equation}
where the WD mass $M_{\rm wd}$ is in units of solar masses. 
At a given $P_{\rm orb}$, the spread in $M_{\rm wd}$ is
expected to be at most $\pm15$\%.

Applying \Eref{McP} we find $M_{\rm wd}=0.20\pm0.03\,M_\odot$ for
\KOI74, which matches the value independently derived in 
\Sref{reanalysis}.  For \KOI81, this becomes $M_{\rm
  wd}=0.25\pm0.03\,M_\odot$, again consistent with a low-mass
WD.\footnote{\citet{taurs99} provide an alternative expression, from
  which we obtain similar values: $M_{\rm wd}=0.24\pm0.02\,M_\odot$
  for \KOI74 and $M_{\rm wd}=0.29\pm0.02\,M_\odot$ for 
  \KOI81.}

Stable mass transfer seems to provide a simple way to produce both
these systems. The match between the WD masses from \Eref{McP}) and
from the light-curve data is encouraging.  Furthermore, the full binary
evolution calculations of \citet*{podsrp02} produce systems with very
similar orbital periods and WD masses (see their Fig.~13).
In \Fref{histories}, we show examples of how the two systems could
have reached their current state, using simple analytic formulae for
orbital evolution (Eq.~(3) of \citealt{rappph09}; see also
\citealt*{podsjh92}).  The curves in \Fref{histories} take initial
conditions \{$M_{1, \rm init}$, $M_{2, \rm init}$, $P_{\rm orb,
  init}$\} of \{$1.6\,M_\odot$, $1.2\,M_\odot$, 19.3\,h\} for \KOI74
and \{$1.8\,M_\odot$, $1.3\,M_\odot$, 31\,h\} for \KOI81.  In both
cases we assume that any mass lost from the system carries away the
specific orbital angular momentum of the binary.  For \KOI74, this
example assumes that the mass transfer is 73\% conservative, and for
\KOI81 it is 91\% conservative.  There are certainly other possible
paths to the current systems, in particular since the above ignores
the possible role of magnetic braking (in evolutionary stages where
either star has a convective envelope).

The currently more massive stars are almost certain to have accreted
significant amounts of matter and are therefore likely to be very
rapidly rotating (see, e.g., \citet{rappph09} and references therein).
Their spin angular momentum could, in principle, have been lost by
magnetic braking.  However, typical A-type stars do not experience
strong magnetic braking \citep[see, e.g.,][]{kawaler88}.
Alternatively, for \KOI74, tides might have been strong enough to slow
the rotation of the star (and slightly widen the orbit).

\subsection{Alternative: Common-envelope Evolution?}

Common envelope evolution seems to provide a much less satisfactory
explanation for these systems than stable mass transfer.  If the
envelopes of the initially more massive stars were ejected during
common envelope evolution, then using the current $P_{\rm orb}$ in
\Eref{McP} would only yield a minimum WD mass.  So, if the
observational data cannot allow more massive WDs than
$\sim\!0.25\,M_\odot$, it seems unlikely that either of these
systems have experienced a significant common-envelope phase.

In addition, any scenario in which the currently more massive stars
have \emph{not} accreted a significant amount of mass seems difficult.
The $>\!2\,M_\odot$ current primary stars in these systems would
produce WDs more massive than $\gtrsim\!0.28\,M_\odot$ if their
envelopes were removed at the end of the main sequence.  If those
initially less massive stars had not gained matter then, at best,
significant fine-tuning would be required for the initially more
massive star to have both evolved off the main sequence and produced
the low-mass WDs observed.

\subsection{Other Expected Compact Hot Companions: Hot Subdwarfs}

Another class of hot compact companion stars that should result in
light curves similar to those in \Fref{lc}, are hot subdwarf (sdB, sdO)
stars, sometimes referred to as extreme horizontal branch stars.  The
calculations of \citet{han+02,han+03} predict that sdB stars should
exist in binaries of these orbital periods about A- and B-type stars
(see \citealt{han+03}, Fig.~15). However, at the temperatures inferred
for \KOI74 and 81, sdB stars would be expected to be considerably more
luminous.  Typical hot subdwarfs properties are anywhere between
$\sim\!20000$ and 40000\,K and $\sim\!3$ and $100\,L_\odot$.  In
addition, the inferred companion masses in \KOI74 is somewhat too low
to have ignited helium, even non-degenerately; masses greater than
$\sim\!0.3\,M_\odot$ are required. Such hot subdwarfs should be found
in current and future planetary transit searches.
 
\section{Summary}

The wonderful photometric precision of {\em Kepler} has allowed us to
measure the radial velocity amplitude of \KOI74.  Combined with the
primary mass inferred from its spectral type and mean density, we
estimated a companion mass of $0.22\,M_\odot$, and argued it was a WD.
We showed that its properties are in very good agreement with
theoretical expectations based on a stable-mass transfer phase inside
a binary system.  A similar evolutionary history is also likely to
have formed the current \KOI81 system.  The many binaries that will be
discovered by photometric surveys should be interesting and useful,
both individually and collectively (see, e.g., \citealt{willkj06} for
a quantitative study), in helping us to increase our understanding of
binary evolution.

Future observations by \emph{Kepler} should allow measurements of the
photometric Doppler effect in more systems, perhaps including systems
such as `ultra-cool white dwarfs,' where the lack of spectral lines
has precluded searches for radial velocity variations. Given long
enough baselines, the photometric analog of the Rossiter-McLaughlin
effect also seems likely to be observed.

\acknowledgements

Coffee-time astro-ph discussions at the Kavli Institute for Astronomy
\& Astrophysics triggered this work; in particular, we thank Yanqin
Wu, Andrew Shannon \& Matthias Gritschneder.  We thank the referee,
Scott Gaudi, for his careful reading and useful comments.  S.A.R.\ and
Ph.P.\ thank J.\ Rowe for sharing his work in advance of submission.
S.J.\ thanks Bill Paxton for the plotting package Tioga.\\

{\it Facilities:} \facility{Kepler}\\

{\em Note added in proof.} After acceptance of this article,
high-resolution spectra were taken for us by E.\ Kirby using HIRES on
the Keck I telescope. From these, we measure projected rotational
velocities of 150 and $225{\rm\,km\,s^{–1}}$ for \KOI74 and \KOI81,
respectively, and infer rotational periods of about 0.6 days for both
sources. For \KOI74, the period is consistent with that inferred from
the weak non-orbital modulation shown in the power spectrum
(\Sref{doppler}).

\bibliography{kepler.bib}
\end{document}